\UseRawInputEncoding
\documentclass[pra,12pt,tightenlines,nofootinbib,floatfix]{revtex4} 
\pdfoutput=1
\usepackage{amsmath}
\usepackage{graphics, graphicx}
\usepackage{epsfig}
\usepackage{ulem}
\usepackage{color}
\usepackage{datetime} 

\begin{document} 





\begin{abstract}
 \centerline{ Brief recollections by the author about  how he interacted with Feynman and was influenced by him.}
\end{abstract}

\title{Recollections of Richard Feynman}

\author{James  Hartle}
\affiliation{Department of Physics, University of California, Santa Barbara, California  93106, USA} {\affiliation{ Santa Fe Institute, \\ 1399 Hyde Park Road,  Santa Fe, New Mexico  87501, USA.}  

\bibliographystyle{unsrt}
\bibliography{references}


\maketitle

\section{Introduction}
\label{intro}

I never wrote a scientific paper with Feynman or worked with him extensively in science.  Despite this, my own work was significantly influenced both by his writings and by  interactions and discussions with him. This paper reports my memories of these interactions and influences. I am relying on my memory  for these knowing  all the risks that go along with that. I  cannot record what I have forgotten in the sixty years since the story began,  I am not writing accurate history that can be backed up by documents.  I am writing  about my recollections of situations and events that occurred long ago. 

\section{ Enter Feynman's Path Integral } 
\label{princeton}  

As an undergraduate physics major at Princeton in the late '50s of the last century  (c.1958-60) I was only taught what might be called the old quantum theory. We learned atomic spectroscopy but nothing of the Schr\"odinger equation. It seems that this was thought to be too advanced for undergraduates.
I could see that this was unsatisfactory and set about learning quantum theory  myself.  

An event that was later important  for me  happened  while still an undergraduate at Princeton. I picked up a discarded reprint from a stack on the floor of Palmer Physical Laboratory. This contained  a reference to Feynman's path integral formulation of quantum mechanics. I immediately read his famous 1948 paper({\it The SpaceTime Approach to Non-Relativistic Quantum Mechanics} ,Rev.Mod.Phys. {\bf 20},367  1948.)  I was struck by the simplicity and beauty of this approach and subsequently worked out as many explicit examples as I could.   Feynman's formulation made clear that quantum theory could be formulated in equivalent but different ways resulting in different pictures of the world \footnote{Possibly the paper I picked up was a reprint of the Proceedings of the Chapel Hill conference in 1957.}.

The importance of the path integral was reinforced when I enrolled  (with permission) in John Wheeler's  graduate mechanics course at Princeton that   started by deriving classical mechanics from quantum mechanics via  Feynman's path integral.
Thus, early on, the Feynman path integral entered my repertoire and never left it.   

\section{First Encounters}
\label{first}   
In 1960 I enrolled in the graduate program at Caltech where Feynman was a professor. 
His presence there was a  significant part of the reason for my choice of schools.  

My first encounters with Feynman were as a graduate student assistant for producing the Feynman Lectures on Physics.  I have written a separate piece  about those experiences  available at arXiv: 2202.07020 which I will not repeat here. 

\section{Christians and Lions}
\label{christ-lions} 
Some idea of what Caltech was like in this period can be drawn from  the character of the theoretical seminars at the time: I  attended them all as did the other theoretical physics graduate students. Even if we could not follow the seminar (as was frequently the case) we could watch the spectacle of the proud and famous being attacked and corrected by the likes of Feynman and Murray Gell-Mann.  Not only were these two  brilliant, informed,  and on top of their fields,  they were fast!  (I once asked Murray:  ``Do people ..... .....sound .... like..... they.....are..... talking ....  to ..... you ..... like....this?''.  He answered ``Why yes!''. ) 

One  time I was sitting in the middle of the room listening to a seminar.  The external speaker had covered multiple blackboards with notes.  Feynman was not present. But he came in during the middle of the talk.  He sat down  beside me, and asked me in a voice easily audible to everyone in the room. ``What is this guy doing?''   I knew a little about the subject, and with the help of the notes on the blackboards began explaining  to Feynman in a voice that could be heard by everyone.  When I was finished I  was exhausted. Feynman was immediately on his feet objecting.  After a bit  of give and take he said to the speaker ``I can give you five reasons why this is wrong  and you can't give me one reason why its right!!  and left the room.

Examples like these taught me to anticipate every objection and be prepared!  

\section{Feynman Lectures on Gravitation}
\label{lectgrav}  
In 1962/63 Feynman gave a course of lectures on gravitation.  The lectures were in  in the afternoons after his lecture to undergraduates  on some of the days when he consulted  at Hughes  Laboratories if memory serves.

 I'm not sure why he gave these lectures. Possibly  he welcomed the challenge of quantizing gravity in the the same way he had worked out the quantum mechanics of electromagnetism. 
 
 The lectures were given  in a small classroom room in the Bridge physics building.  The room could at best hold about 20 people. Remarkably, given Feynman's standing in physics, that was all that was necessary to accommodate those interested. 
 Of course, as a card carrying  worker in gravitational physics I aimed to attend every lecture. 
 
 One day, because of a flat on my roommate's car, I was 12 minutes late for a lecture. Feynman looked at me in the first row and said ``You! You're late!'' I started to explain but Feynman cut me short saying.  ``Ok,  I'm going to start the lecture all over again just for you, but I am going to go really fast!'' Of course Feynman lecturing at top  speed was impossible to understand,  but I later put the missed pieces together.  I would like to think that Feynman saw something in me that made it worthwhile for him  make  sure I got the point. 
 
 Another incident in these lectures says something about Feynman.  Before one lecture someone had written in a bold hand on an otherwise pristine black board. ``Man does not live by bread alone.''  Feynman came in, saw it, and completed the phrase so it read 
 ``Man does not live by bread alone save by every word that proceeds from the mouth of Feynman.''  I wouldn't have guessed, then or now, that he could or would quote  Mathew(4-4).
 
  A call went out for students to write the lecture notes. I was keen to do this and raced to Feynman's office to volunteer. But my teacher, Gell-Mann was ahead of me. He had anticipated this and  called Feynman and told him not to allow me to do it. Ñ I had too much important work to do. The notes were written up by Fernando Morinigo and W. G. Wagner,  and are still available as
R.P. Feynman,   ({\it Lectures on Gravitation, notes by F. Morinigo and W.F Wagner, (Frontiers in Physics,  CRC Press, 2002).} (For much more on the Lectures see John Presskill's introduction to the notes.)

\section{Impact of the Feynman Lectures on Gravitation on My Scientific Work.}
\label{impact-lectgrav} 
 Feynman did not begin his lectures on gravity by  assuming Newton's several  hundred years old law of gravity or Einstein's relativistic theory of gravity --- general relativity.  He aimed at seeing whether these could be {\it explained} by something already known in physics. He wanted to find out {\it why} they existed. He tried out things like the solar wind, a field coupled to baryon number, etc, etc all ruled out experimentally. His lectures were as interesting for what didn't work as what did.

 \subsection{Neutrino Exchange} 
 \label{neutrino} 
 The exchange of zero mass neutrinos would seem to result in a long-range force and Feynman considered that at length.   But according to the 4-point theory of the weak interactions used  at the time two neutrinos would have to be exchanged.  Simple dimensional considerations showed that would yield an inverse $r^5$ force. --- not gravity. I later confirmed all this with a quantum field theory calculation as did other people, e.g.   
(G. Feinberg and J. Sucher, Phys. Rev. {\bf 166} 1638, 1968.))

 \subsection{Enter Cosmology}
 \label{cosmo} 
 At this impasse Feynman had a marvelous idea.  Exchange only a single neutrino between the two particles of interest resulting in an inverse squared force between them. Exchange the second neutrino with all the other particles in the Universe resulting locally in an inverse squared force between pairs of particles --- gravity
 
 \subsection{Long Range Neutrino Forces}
 \label{neutrino_forces}
 I don't recall that Feynman did anything to follow his great idea up. But I did.  I reasoned that long range neutrino forces might not be an explanation of gravity but could be important for cosmology and black holes anyway. I wrote several papers on this subject:
 
  J. B. Hartle,  {\it Long Range Weak Forces and Cosmology.} {\sl Phys. Rev. D} {\bf 1 }, 394-397, (1970).

J.B. Hartle {\it  Long Range Neutrino Forces Exerted by Kerr Black Holes.}, {\sl Phys. Rev. D}, {\bf} 3, 2938- 2940, 1971).

J.B. Hartle,  {\it Can a Schwarzschild Black Hole Exert a Long Range Neutrino Force?}, in Magic Without Magic (eds. J. A. Wheeler, J. Klauder), W. H. Freeman, Co., San Francisco, (1972).

\subsection{Deriving the Einstein Equation}
\label{DerivEE}   
In his Lectures  on Gravitation  Feynman did not posit the Einstein equation. Rather he derived it from the simple principle that gravity is sourced by mass-energy. Applying this principle step-by step, in successive orders of perturbation theory, and summing all the orders up he derived the Einstein equation. ---  curvature tensors and all.  Along the way new technical tools useful for perturbation theory elsewhere were developed (e.g ghosts).   I thought all this an amazing tour de force at the time.Similar calculations were carried out  by others e.g.Steven Weinberg,  {\it  Photons and Gravitons in Perturbation Theory: Derivation of Maxwell's and Einstein's Equations} 
Phys. Rev. {\bf 138,} B988, (1965). 

\subsection{ Deriving  Born's  Rule} 
\label{deriving BR} 
My interactions with Feynman did not  stop with classical general relativity.  They continued on to quantum mechanics. 
By 1968 I had developed a new interest ---  finding a better understanding of quantum mechanics  motivated by much seeming confusion as to what it means.  I introduced many ideas in my first paper  on this subject:
(J.B. Hartle, {\it  Quantum Mechanics of Individual Systems} ,{\sl  Am. J. Phys}  {\bf 36}, 704-712, (1968).) I was especially proud of my derivation of Born's rule from simple assumptions in what today would be called the quantum mechanics of closed systems 
like the  Universe.  [There have been a great many derivations since e.g.
(J. B. Hartle, {\it What do We Learn by Deriving BornÕs Rule? }, arXiv:2107.02297)]

\section{A Theoretical Seminar at Caltech}
 I thought that my derivation of Born's Rule was my best accomplishment in physics up to that time. Naturally I wanted to show it to Feynman. My chance came with an invitation to give a theoretical seminar at Caltech.  The day came and Feynman was in the audience! 
 
 My  seminar did not go well.  I started my talk by writing one equation on the blackboard and was immediately attacked. The attacks came not so much  from Feynman himself but from the pilot fish that accompanied him. My guess is that it was difficult for Feynman to believe that there was something new in quantum mechanics that he did not already know.  That one equation was the only one I was to write.
 
 His final pronouncement at the end of the hour was:  ``sounds good, sounds good''.  I took this to mean ``I don't believe it but I don't have any supportable objections''.  I  could live with that. 
 
 Examples of treatment like this taught me to anticipate every objection and be prepared! 

\section{A Physics Colloquium at Caltech}
\label{citcolloq} 
A physics  colloquium at Cal Tech. provided an another occasion for me to describe my work with Feynman present.  I decided to include my derivation of Born's rule. The approximately 300 seat lecture hall was full. Feynman sat in the middle of the third row --- eye to eye with the standing speaker as was his wont.  At one point in my. talk I made an assertion.   Feynman in a voice easily heard by everyone said ``Which never happens''.  I went on with my lecture and demonstrated with unit probability that ``it always happens''.  I  then looked straight at Feynman. There was a very long pause but then he began nodding up and down ---``Yes!''. The apparent relief on my face led the entire audience to laugh and cheer (sympathetically) for some minutes. I was congratulated by many afterwards.

\section{A Missed Opportunity} 
\label{missed} 
I have always regretted that, despite trying,  I did not get to explain to Feynman the resolution  of an issue he didn't resolve that  he describes in his 1948 classic. There he says: 
`\it Suppose a measurement is made of a particle at one time  which is capable only of determining that it's path lies somewhere within  a region of space $\cal R$, `The measurement is to be what we might call an Òideal measurement. We suppose that no further details could be obtained from the same measurement without further disturbance to the system. I have not been able to find a precise definition'}. 

Feynman's issue is resolved in the modern formulation of quantum mechanics called  Decoherent or Consistent histories quantum theory  (DH) .  (See, e.g. QuantumMechanics in the Light of Quantum Cosmology, M.Gell-Mann and J.B. Hartle, arXiv/1803.04605.)  The  question of whether a particle's path lies in a region $\cal R$  at some time defines a set of two alternative coarse-grained {\it histories } whose probabilities can be computed DH    given the quantum state of the particle.e.g. as in {\it  Prediction in Quantum Cosmology}, in Gravitation in Astrophysics (ed. by B. Carter and J. Hartle), Plenum Press, New York, 329-360, (1987), arXiv:2108.00494.  

I asked for a chance to explain this to him, but he was already seriously ill and declined. He died not long after. 

\section{Feynman and the Quantum Mechanics  of the Universe} 
\label{everett} 
I don't think I ever had the opportunity to discuss, or listen to a discussion, of whatever views on quantum mechanics Feynman might have. However, K[p Thorne  told me that on one occasion he had asked Feynman what interpretation of quantum mechanics he favored.   Feynman  responded that Òwhen pushed to the wallÓ he felt forced to fall back on the Everett interpretation.  He said he doesnÕt like it, but he had nothing better. Murray Gell-Mann told me that  in '63-'64 he, Feynman, and Felix Villars had worked together on formulating quantum mechanics with a result that was close to Everett. This has also been suggested by some current authors\, e.g (H.D. Zeh, {Feynman's Interpretation of Quantum Theory}Eur. Phys. J. H36, 147 (2011) ,  arXiv:0804.3348v6)

\section{Summing Up} 
\label{summing} 
I had only a few personal interactions with Feynman but they were almost all consequential for my work.  I had only a few chances  to explain my work. In  the end, I don't know what he thought about it.   However, I take some satisfaction in the thought that even ill and headed for death he would make some time to listen to me about what I was doing.

\section{Conclusion} 
Feynman bulks large in my recollections because of the   the  ideas and  opportunities that his work provided for me --- ideas that I followed up, elaborated, and applied in different physical situations.  I had only a few chances to explain my work to him but the comments that resulted were always  relevant.  I remember him  vividly. I hope these modest recollections show something of the man and how he worked to make the great advances in science he  did.

\section{Acknowledgements} 
\vskip .1in 
\noindent{\bf Acknowledgments:}    Thanks  are due to the NSF for supporting its preparation under grant PHY-18-8018105 and to Mary Jo Hartle for proofreading it more than once.

 \end{document}